\documentclass[aps,prd,twocolumn,nofootinbib,groupedaddress,amsfonts,floatfix]{revtex4}

\usepackage{graphicx,amsmath,amssymb,amstext}
\usepackage{amssymb,amsbsy,amsfonts,amsthm}

\usepackage{color}
\usepackage{ifthen}
\newboolean{editorial}
\setboolean{editorial}{true}
\newcommand{\editorial}[2]{\ifthenelse{\boolean{editorial}}{\textcolor{red}{[\textsf{\textbf{{#1}}}: }\textcolor{blue}{\textsf{{#2}}}\textcolor{red}{]}}{}}

\begin{document}

\title{Gauge Field Preheating at the End of Inflation}
\author{J. Tate Deskins${}^1$}
\author{John T. Giblin, Jr${}^{1,2}$}
\author{Robert R. Caldwell${}^3$}
\affiliation{${}^1$Department of Physics, Kenyon College, Gambier, OH 43022, USA}
\affiliation{${}^2$Department of Physics, Case Western Reserve University, Cleveland, OH 44106, USA}
\affiliation{${}^3$Department of Physics and Astronomy, Dartmouth College, Hanover, NH 03755, USA}

\begin{abstract}
Here we consider the possibility of preheating the Universe via the parametric amplification of a massless, $U(1)$ abelian gauge field.  We assume that the gauge field is coupled to the inflaton via a conformal factor with one free parameter.  We present the results of high-resolution three-dimensional simulations of this model and show this mechanism efficiently preheats the Universe to a radiation-dominated final state.  
\end{abstract}

\maketitle

\section{Introduction}

Apparently it is easy to inflate the Universe. There is no lack of theoretical models that predict a deSitter-like expansion epoch in the early Universe \cite{Kachru:2003sx,Linde:2005ht,Burgess:2007pz,Baumann:2009ds}. The present-day cosmic expansion appears to be inflating as well \cite{Riess:1998cb,Perlmutter:1998np}. Yet, reheating the cold Universe at the end of primordial inflation is more of a challenge. In many models, inflation must be followed by a long phase in which the inflaton oscillates at the bottom of its potential, slowly transferring its energy to the rest of the Universe, to which it is only weakly coupled. Although the old theory of reheating \cite{Dolgov:1982th,Abbott:1982hn,Bassett:2005xm} works, there has always been a desire to reheat the Universe more efficiently, hence, achieving a large reheating temperature. One popular way to accomplish this is {\sl pre}heating, whereby classical field effects \cite{Traschen:1990sw,Kofman:1994rk,GarciaBellido:1997wm,Khlebnikov:1997di,Greene:1997ge,Parry:1998pn,Bassett:1998wg,GarciaBellido:1998gm,Easther:1999ws,Liddle:1999hq,Finelli:2001db,Bassett:2005xm,Podolsky:2005bw} quickly and efficiently transfer the inflaton energy to matter fields via three- or four-leg interactions despite weak couplings and small decay rates.

In a typical preheating scenario, the inflaton $\phi$ couples $g^2\phi^2\chi^2$ to another scalar field $\chi$ whose mass is assumed to be small compared to the expansion rate. The oscillation of the inflaton amplifies particular modes of $\chi$, akin to the way a child pumping  her legs amplifies the harmonic motion of a playground swing. This driven instability creates strong inhomogeneities in both $\chi$ and $\phi$, whereafter it is generally assumed that the $\chi$ and $\phi$ particles decay into the ultra-relativistic species of the Standard Model. This is the largest success of preheating; it is the mechanism through which the Universe heats up to create the Hot Big Bang. Unfortunately, most models of preheating, the Universe is still looks matter-dominated at the end of the resonance period before thermalization takes place. Moreover, these same models leave us with few observational signatures. It is likely that the only observable consequence is in the form of a stochastic gravitational wave background \cite{Khlebnikov:1997di,GarciaBellido:1998gm,Easther:2006vd,Easther:2007vj,Dufaux:2007pt,Easther:2006gt,Dufaux:2008dn,Price:2008hq} whose signal will be very difficult to detect unless inflation occurred at a very low scale.

In this manuscript we study the intriguing possibility that the fields that undergo preheating are massless gauge fields. We use electromagnetism with a dilaton-like coupling to the inflaton that breaks conformal invariance as the basis of our model, though more complicated gauge fields may also be considered, e.g. \cite{ArmendarizPicon:2007iv}. Our model has long been proposed as a mechanism for generating primordial magnetic fields. (See e.g. Refs.~\cite{Turner:1987bw,Ratra:1991bn,Lemoine:1995vj}, and Refs.~\cite{Bamba:2003av,Martin:2007ue,Demozzi:2009fu,Kanno:2009ei,Subramanian:2009fu,Durrer:2013pga} for more recent articles and review.) New impetus to study this model comes from general interest in a broader range of interactions of the inflaton, particularly with gauge fields, motivated in part by recent work on the effective theory of inflation \cite{Cheung:2007st,Weinberg:2008hq} and also new models of inflation that directly utilize gauge fields \cite{Maleknejad:2011jw,Maleknejad:2012fw,Adshead:2012kp,Dimastrogiovanni:2012st}. Furthermore, it has recently been demonstrated that fluctuations of the inflaton are correlated with the large-scale magnetic field as a result of the coupling in our model \cite{Caldwell:2011ra,Motta:2012rn,Jain:2012ga}. Hence, a residual cross correlation in the large scale pattern of inhomogeneities in the sky may provide a signal as to the mechanism of preheating.  

In our model, a single scalar degree of freedom, $\phi$, is responsible for inflation and is coupled to electromagnetism via
\begin{equation}
S=\int d^{4}x\sqrt{-g}\left( -\frac{ W(\phi)}{4} F_{\mu\nu}F^{\mu\nu}-\frac12\partial_{\mu}\phi\partial^{\mu}\phi-V(\phi)\right)
\label{eqn:action}
\end{equation}
where $W(\phi)$ and $V(\phi)$ are scalar functions and represent the coupling strength and the inflationary potential respectively.  Although we identify the $U(1)$ gauge field with the usual electromagnetic field, there is no explicit need that this be the case. We use a standard conformal metric for an expanding Friedmann-Lema\"itre-Robertson-Walker (FLRW) space-time
\begin{equation}
ds^{2}=a^2(\tau)(-d\tau^{2}+d \vec x^2).
\end{equation}
The scalar field, $\phi$, is subject to a quadratic potential parameterized by a mass scale $m_\phi$,
\begin{equation}
V(\phi)=\frac12 m_\phi^{2}\phi^{2}.
\end{equation}
For the majority of the work presented here, we fix $m_\phi = 10^{-6}\,m_{\rm pl}$, where we define the Planck mass in terms of the gravitational constant as $m_{\rm pl} = 1/\sqrt{G}$.  The coupling function $W(\phi)$ has the form 
\begin{equation}
W(\phi)=e^{\phi/M},
\label{eqn:concouple}
\end{equation} 
where $M$ is a free parameter which will ultimately take a value near the geometric mean of $m_\phi$ and $m_{\rm pl}$.  As the scalar field decays, the coupling goes to unity and standard electromagnetism is recovered.

The equations of motion of the coupled scalar-vector system are simply $\nabla_\mu(W F^{\mu\nu})=0$ and $\Box\phi = V' + \tfrac{1}{4} W' F^2$. Hence, the coupling function $W$ acts like a source of charge and current density for the electromagnetic fields, which in turn contribute an effective mass to the scalar field. To simplify the system of equations, we impose the Lorenz gauge condition
\begin{equation}
\partial_\tau \Phi+ \vec\nabla \cdot \vec A = 0
\label{gaugecondit}
\end{equation}
where $A_\mu = (A_\tau, \vec A)$ and $\Phi = -A_\tau$ is the scalar potential. The gradient operator is evaluated on the comoving, Cartesian space. Consequently, we find the equation of motion for the electromagnetic potentials
\begin{equation}
\left(\partial_\tau^2  - \nabla^2\right)A_\mu = J_\mu  
\label{eqn:eomamu}
\end{equation}
where $J_\mu = (-\rho_{\rm eff}, \vec J_{\rm eff})$ and
\begin{eqnarray}
\rho_{\rm eff} &=& \frac{1}{M}\vec\nabla\phi \cdot(\partial_\tau \vec A  - \vec\nabla A_\tau) 
\label{eqn:rhoeff}\\
\vec J_{\rm eff} &=& -\frac{1}{M}\left[  \partial_\tau \phi ( \partial_\tau \vec A - \vec\nabla A_\tau ) +
\vec \nabla \phi \times (\vec\nabla \times \vec A)\right].
\label{eqn:jeff}
\end{eqnarray}
These effective source terms are conserved, such that $ \partial_\tau\rho_{\rm eff} + \vec\nabla\cdot \vec J_{\rm eff} = 0$. The equation of motion for $\phi$ is
\begin{eqnarray}
&&\partial_\tau^2 \phi + 2 {\cal H} \partial_\tau \phi - \nabla^2\phi + a^2 V_{,\phi} \cr
&&=\frac{W_{,\phi}}{2 a^2}
\left[ (\partial_\tau \vec A   - \vec\nabla A_\tau)^2 - (\vec\nabla\times\vec A)^2\right]
\label{eqn:eomphi}
\end{eqnarray}
where ${\cal H} = d \ln a/d\tau$.

The energy density in the scalar field is
\begin{equation}
\rho_{\phi}= \frac{1}{2 a^2} \left[  \left( \partial_\tau\phi \right)^{2}+ (\nabla\phi)^{2} \right] + V,
\end{equation}
whereas the energy density in the vector field is
\begin{equation}
\rho_{\rm EM} = \frac{W}{2 a^4} \left[
( \partial_\tau \vec A  - \vec\nabla A_\tau)^2 + (\vec\nabla\times\vec A)^2 \right],
\end{equation}
which is the standard energy density for electricity and magnetism enhanced by our conformal coupling, $W(\phi)$.  We also solve for the cosmic expansion, via
\begin{equation}
{\cal H}^2 = \frac{8 \pi a^2}{3 m_{\rm pl}^2} \left( \rho_\phi + \rho_{\rm EM} \right),
\end{equation}
to close the system of equations.

\section {Numerical Simulations}  

The art of three-dimensional lattice simulations is now mature.  A family of numerical codes, first {\sc LatticeEasy} \cite{Felder:2000hq} and then {\sc Defrost} \cite{Frolov:2008hy}, {\sc PSpectRE} \cite{Easther:2010qz}, and {\sc HLattice} \cite{Huang:2011gf}, have been used to explore linear and non-linear dynamics in expanding space-times.  To improve efficiency without reducing accuracy, these codes all employ clever rescalings that reduce the Klein-Gordon equation to a phase-separable equation, and hence, they can use symplectic integrators which require a fraction of the physical memory of other methods.  Here, we push the problem beyond the boundaries of these numerical techniques, since Eqs.~(\ref{eqn:eomamu}) and (\ref{eqn:eomphi}) are not phase separable under the same coordinate transformation.  The field values as well as the time derivatives of the field must be known at the same time. To deal with this we employ {\sc Grid and Bubble Evolver} ({\sc GABE}) \cite{GABE}, which uses a second-order Runge-Kutta method of integration.  This method requires about twice as much physical memory (and longer run-times) than its symplectic cousins, but is more versatile and able to adapt to our model.

We begin our simulations at the point when inflation ends.  The homogeneous modes of the field and its derivative, $\phi_0 = 0.2 \,m_{\rm pl}$ and $\partial_\tau{\phi}\approx 0.14 \,m_{\phi}m_{\rm pl}$, are determined by the inflationary dynamics. We set the fluctuations of the field to be consistent with the Bunch-Davies vacuum,
\begin{equation}
\langle | \tilde\phi(k) |^2\rangle = \left( 2a \sqrt{k^2 + m_\phi^2}\right)^{-1},
\end{equation}
and we use the Fourier convention 
\begin{equation}
\tilde{f} = (2\pi)^{-3/2} \int d^3x\,e^{-i\vec{k}\cdot\vec{x}} f(x).
\end{equation} 
The majority (usually all) of our modes are smaller than the horizon when the simulations begin, and so we ignore any effects that might arise for modes larger than the horizon \cite{Frolov:2008hy}. The initial fluctuations for $A_{i}$ are set differently from the fluctuations for $\phi$--we must scale the fluctuations by $1/\sqrt{W(\phi_{0})}$ as in \cite{Caldwell:2011ra,Motta:2012rn,Maleknejad:2012fw}. The power spectrum for each component of the $A_{i}$ are given by
 \begin{eqnarray}
\nonumber \langle |\tilde A_{i}(k)|^2\rangle 
&=& \langle \tilde A_i(k) \tilde A_j(k^\prime) \delta^{ij} \delta (k-k^\prime)\rangle  \\ 
&=& (2a k W(\phi_{0}) )^{-1}.
\end{eqnarray}
For all of the results presented here we use a $256^3$ lattice and an initial box size of $L_0 = 20 \,m_{\phi}^{-1}$.  We initialize all simulations at $\tau=0$ and $a=1$. The computational method sets initial conditions in momentum space, drawing the magnitude and phase of modes from a distribution so that the configuration space fields have the correct statistics.  The gauge condition is imposed as an additional constraint on the initial conditions. We set $A_\tau = 0$, and then assign random initial conditions (in momentum space) to the $i,\,j,\,k$ components of the gauge field that are consistent with $\vec k \cdot\vec A = 0$. In principle, satisfying the gauge condition on the initial slice in momentum space should satisfy the gauge condition in configuration space.  In practice, however, the existence of high-frequency modes makes evaluating finite-spatial derivatives inaccurate.  To solve this problem, we apply a window function to our initial conditions,
\begin{equation}
\label{windowf}
F(k) = \frac{1}{2} \left[1-\tanh\left(s (k - k_*)\right)\right]
\end{equation} 
where $k_*$ sets the scale of the cutoff and the parameter $s$ dictates the smoothness of the cutoff.  For the results presented in the following section we chose $k_* = k_{\rm nq}/4$ and 
\begin{equation}
s=\frac12 \frac{L_0}{2\pi}\,m_\phi^{-1} \approx 1.56\,m_\phi^{-1}.
\end{equation}
The scale $k_{\rm nq}$ is the Nyquist frequency for our box,
\begin{equation}
k_{\rm nq} = 256\sqrt{3}\frac{2\pi}{L_0} \approx 140 m_\phi, 
\end{equation}
and is the largest wave vector we can resolve on our box.  Since the highest frequency modes have a characteristic comoving period of 
\begin{equation}
T_{\rm min} =  \frac{2\pi}{\omega}= \frac{2\pi}{k}=\frac{L_0}{N},
\end{equation}
we set the dimensionless conformal timestep, $h= d\tau \,m_{\phi}$, so that we have sufficient temporal resolution for this highest frequency mode.  In other words, we want to make sure that the ratio of the shortest period to the timestep,
\begin{equation}
\frac{T_{\rm min}}{d\tau} =  \frac{L_0}{ N } \frac{1}{h} ,
\end{equation}
is larger than 10 or so, thus we have at least ten slices over the course of one oscillation. To achieve this, we set $h = 0.005$, which ensures that $T_{\rm min}/d\tau \approx 15$.  We terminate the simulation when $a\approx 15$ or so, when most of the resonant behavior in which we are interested ceases to occur.

\begin{figure}[htb] 
\centering
\includegraphics[width=0.99\columnwidth]{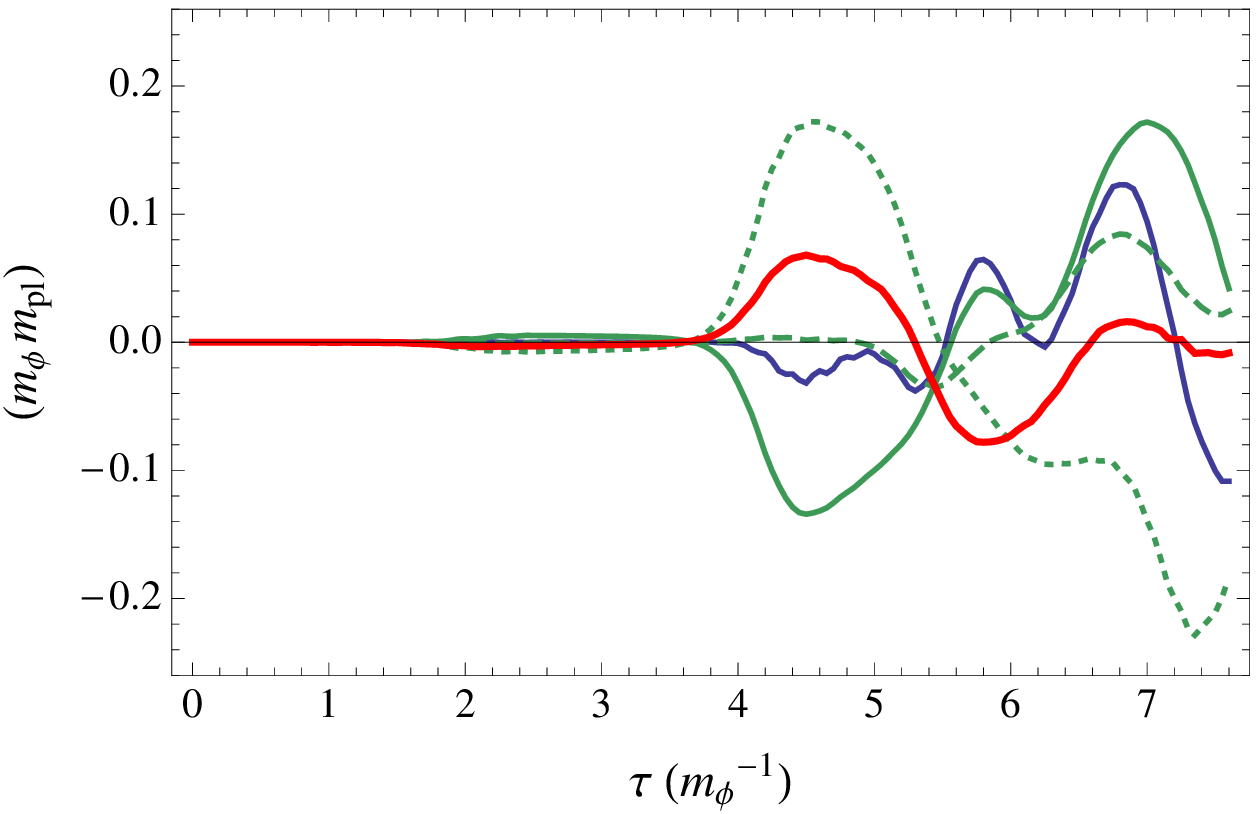}
\includegraphics[width=0.99\columnwidth]{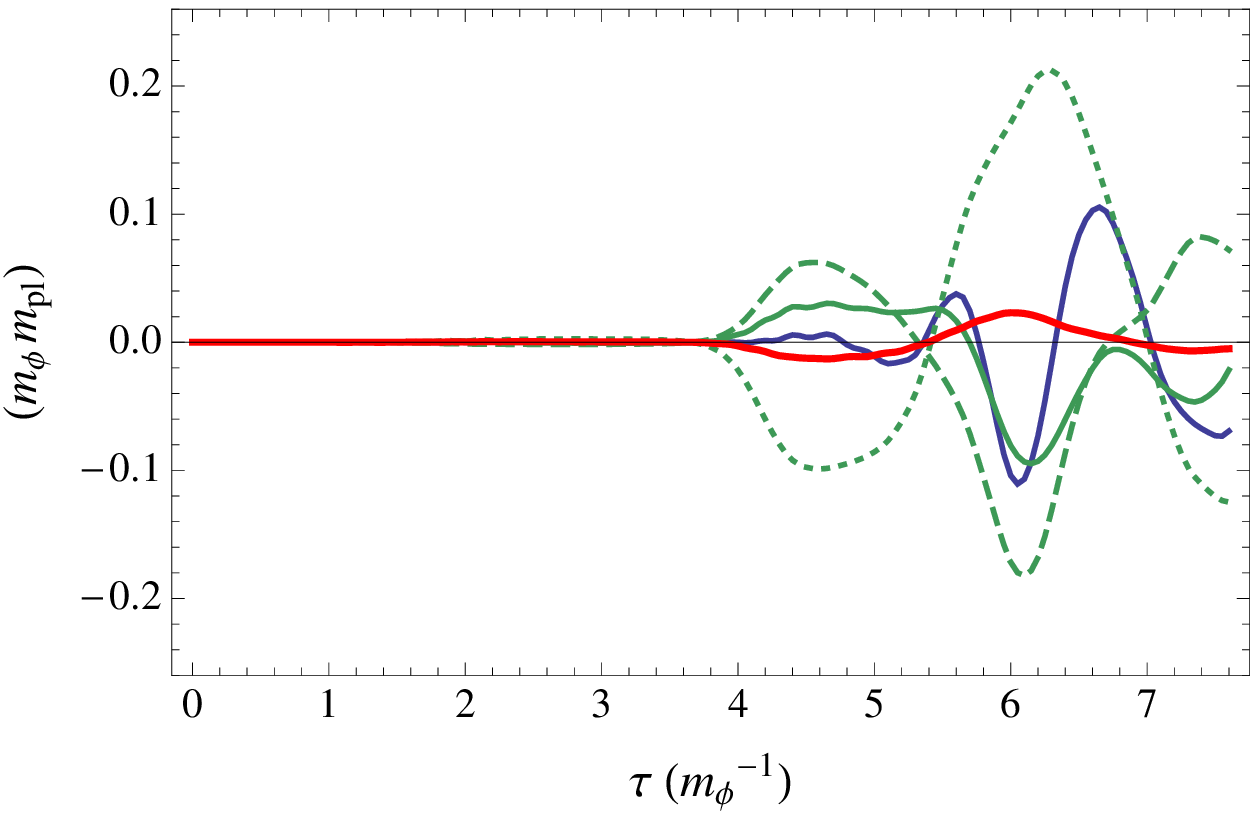}
\includegraphics[width=0.99\columnwidth]{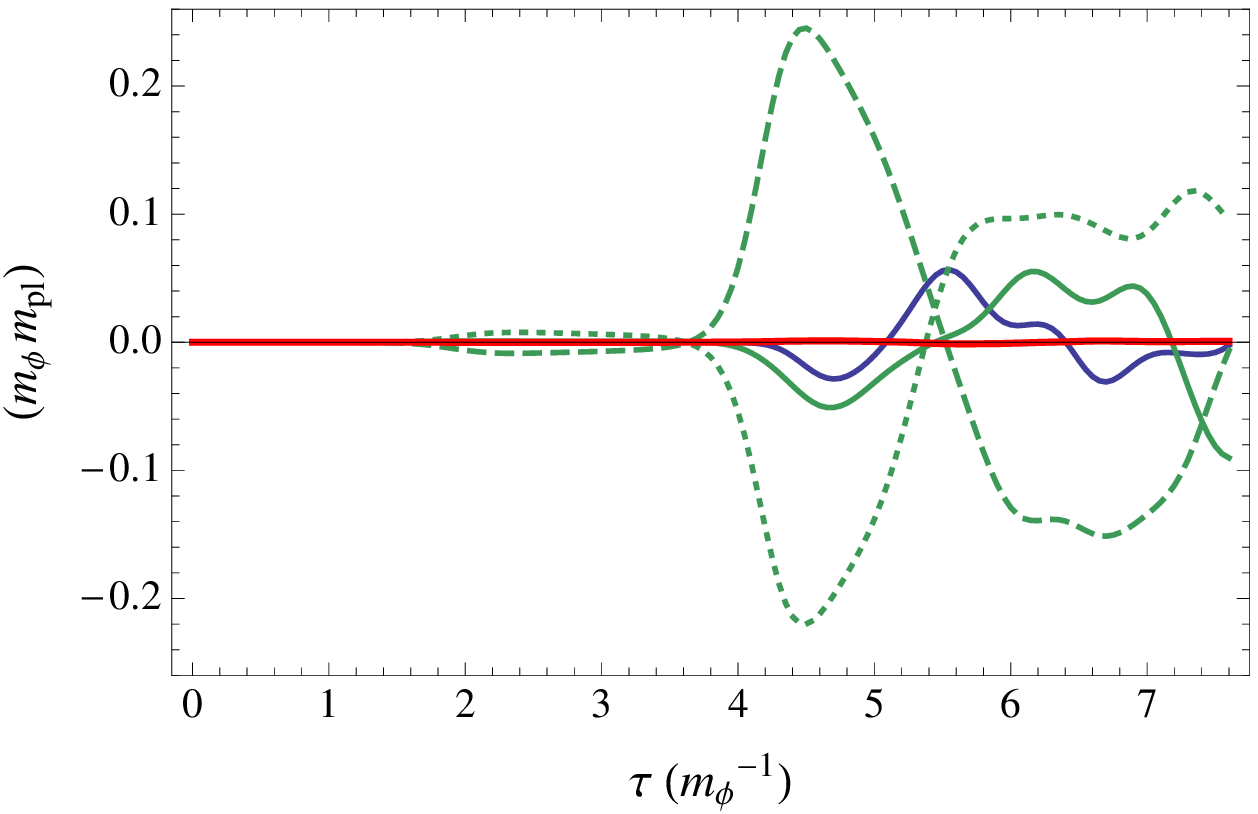}
\caption{The four terms that go into calculating the gauge constraint, $-\partial_\tau A_\tau + \vec\nabla\cdot\vec A$ (red, solid), consisting of $\partial_\tau{A_\tau}$ (blue, solid), $\partial_x A_x$ (green, solid), $\partial_y A_y$ (green, dashed) and $\partial_z A_z$ (green, dotted), for three different simulations. The top panel shows the behavior in the absence of a cutoff function, the middle panel shows $k_* = k_{\rm nq}/2$ and the bottom panel shows $k_* = k_{\rm nq}/4$.  In all cases,  $m_{\phi}=10^{-6}\,m_{\rm pl}$ and $M=0.016\,m_{\rm pl}$, and the field strengths are calculated at an arbitrary point inside the box.}
\label{fig:gaugecondits}
\end{figure}

The introduction of the window function, Eq.~(\ref{windowf}), and its associated parameters is precautionary, too.  Fig.~\ref{fig:gaugecondits} shows how well the gauge condition is satisfied at a particular point in our box over the course of the simulation. We see that as we decrease the cutoff frequency, $k_*$, the gauge condition is more accurately satisfied.  
Satisfaction of the gauge condition is not sensitive to the value of the smoothing parameter for the range of values we tested, $0.1\lesssim s\lesssim 1$. Although the existence of a window function damps out the higher frequency modes, and hence, changes the average initial energy in the gauge field (as can be seen in Fig.~\ref{fig:cutoffedratio}), there is little effect on the physics; the evolution of the box, $a(\tau)$, the existence of resonant amplification of modes, and the final state of the simulation are largely insensitive to $k_{*}$, as we will show in the following section. 
 
\section{Results}
\label{results}

The inflaton is nearly homogenous and preparing to enter a phase of coherent oscillations at the beginning of the simulations.  It is usually this period of coherent oscillations that witnesses the resonance typical of preheating. At this early stage, $|\nabla\phi| \ll |\partial_\tau \phi|$ so that the source term $\rho_{\rm eff}$ is negligible, which is consistent with the initial condition $A_\tau = 0$. The current density at this time is approximately $\vec J_{\rm eff} \simeq -\partial_\tau \phi \partial_\tau \vec A/M$. Although the source term is non-linear, we may approximate $\phi (\vec{x},\tau) \rightarrow \phi (\tau) \approx \bar\phi \cos(m_\phi \tau)$ at early times. In this case, the mode equations for $\vec A$ are
\begin{equation}
\partial^2_\tau \vec A + k^2 \vec A \approx  \frac{\bar\phi m_\phi}{M} \sin(m_\phi \tau) \partial_\tau \vec A.
\label{eqn:eomamuapp}
\end{equation}
The mode equation does not take the form of a Mathieu equation, as commonly seen in preheating, but the oscillatory inflaton very clearly pumps energy into the vector field when the coefficient of $\partial_\tau \vec A$ is negative. Hence, $\vec A$ grows and sources fluctuations of $\phi$, which feed back into the sources $\rho_{\rm eff}$ and $\vec J_{\rm eff}$.

\begin{figure}[htbp] 
   \centering
\includegraphics[width=.99\columnwidth]{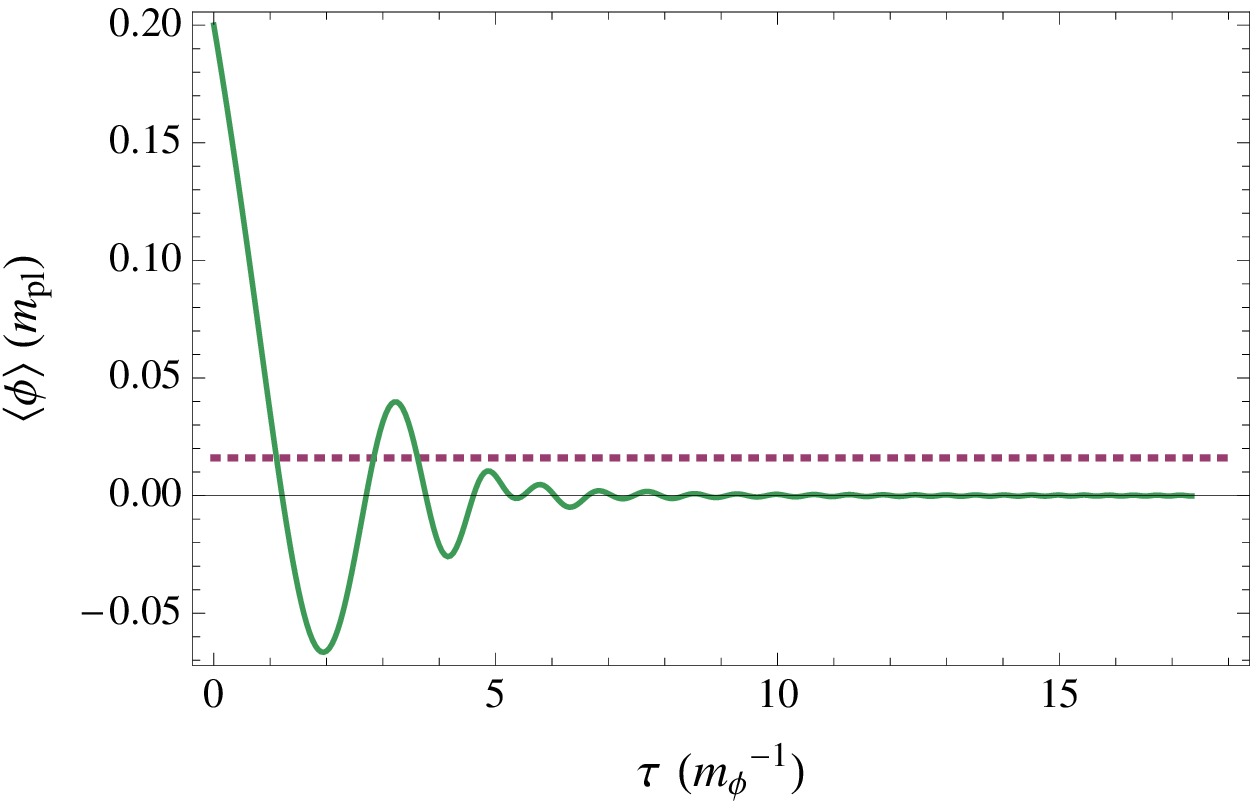}
\includegraphics[width=.99\columnwidth]{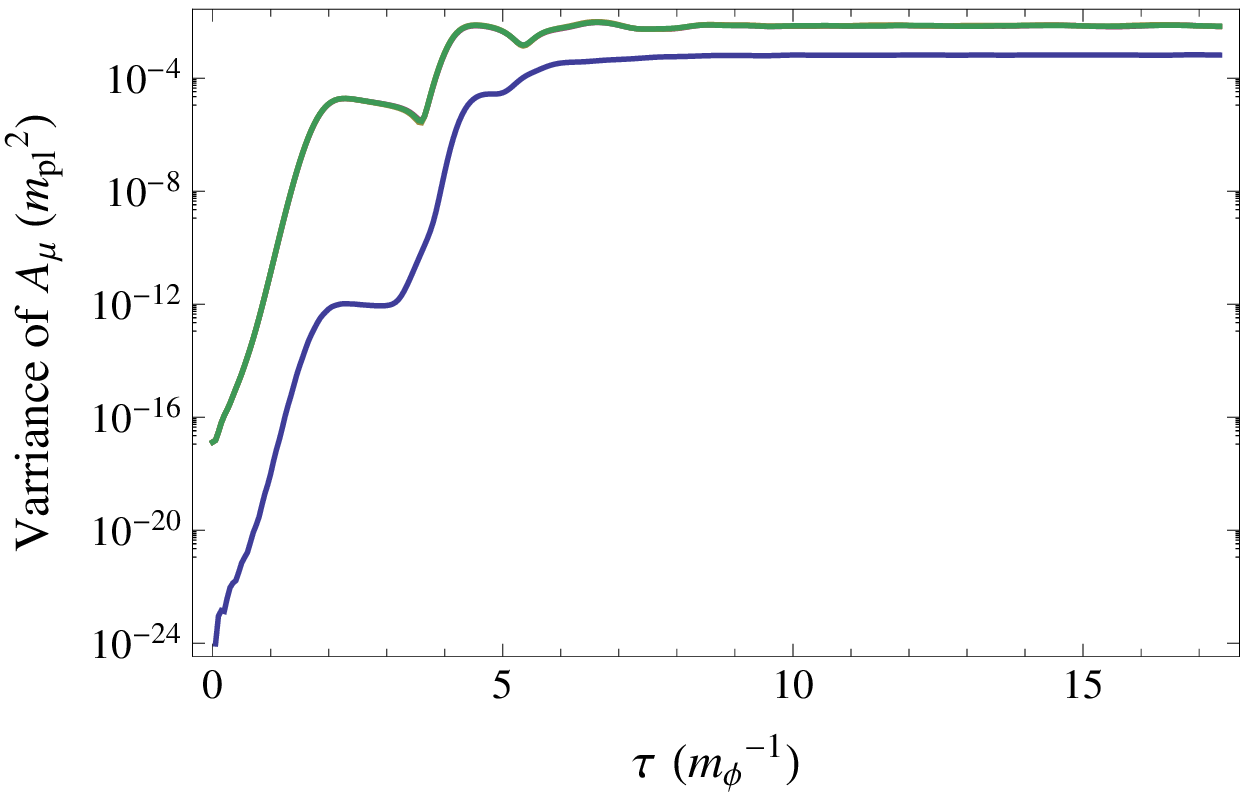}
\includegraphics[width=.99\columnwidth]{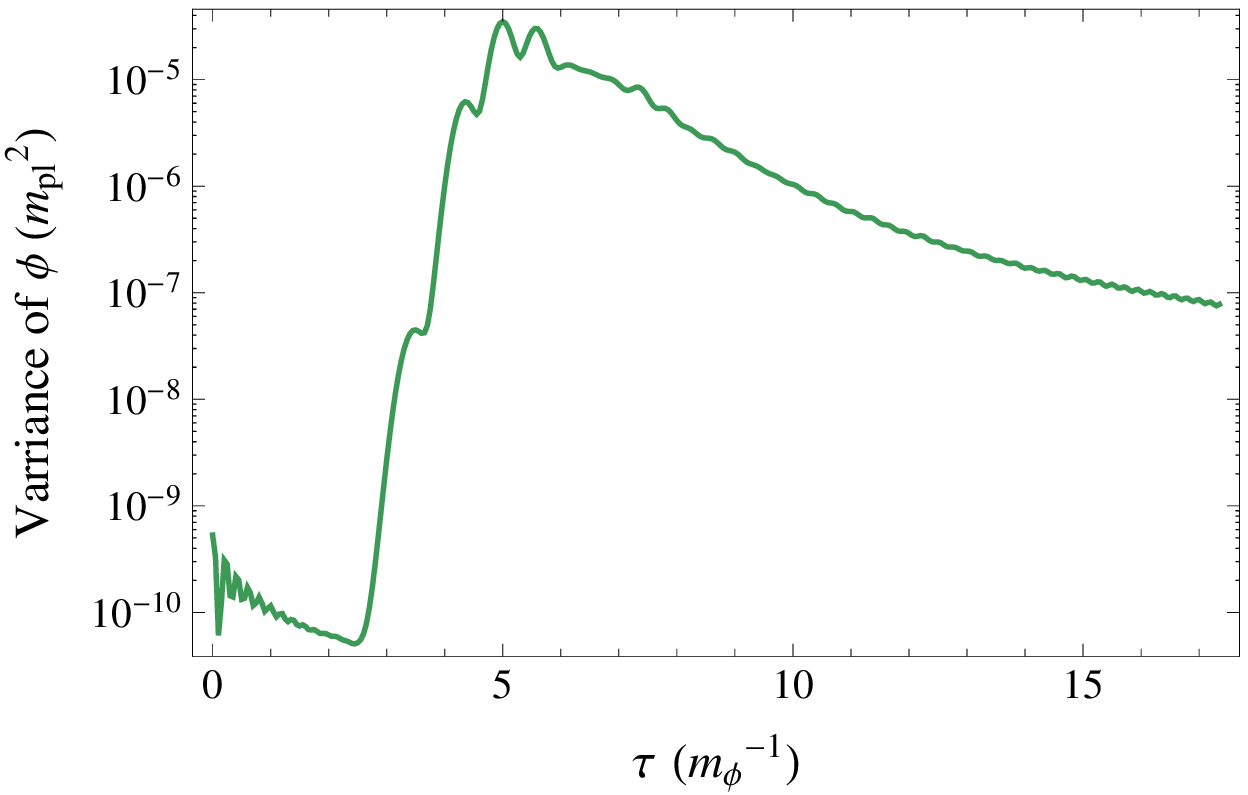}
\caption{The mean value of $\phi$ over the box (green, solid) with reference lines at $\phi=0$ (black, solid) and $0.016\,m_{\rm pl}$ (black, dashed) (top), the variance of the gauge fields $A_{\tau}$ (blue, solid) and $\vec A$ (green, solid) (middle), and the variance of the inflaton $\phi$ (bottom).}
   \label{fig:fields}
\end{figure}

We expect the source term in Eq.~(\ref{eqn:eomamuapp}) to become less important as the ratio of $\phi/M$ decreases. The inflaton can be said to decouple from the gauge field when $\phi \ll M$ with negligible oscillations. This expectation is realized in our simulations, as can be seen in Fig.~\ref{fig:fields}.  First, this figure shows how the amplification of the modes of the gauge field (parameterized by an increase in the variance of $A_\mu$) is strong when $\phi/M$ is large.  Second, the variance of the inflaton also increases during this resonance period, thereby amplifying high frequency modes. Fig.~\ref{fig:fields} illustrates that when $\phi$ is comparable to $M$ this mechanism begins to shut off and the resonant production of energy in the gauge field ceases.  For the case of $M=0.016\,m_{\rm pl}$, this occurs around $\tau\sim 5\, m_{\phi}^{-1}$.

It is interesting to note that the time component of the gauge field, $A_\tau$ (blue curve in the middle plot of Fig.~\ref{fig:fields}), has a lower variance over the duration of the run since $A_\tau$ is initialized to identically zero at the beginning of the simulation.  Even still, the variance of all components of $A_{\mu}$ grows by about seventeen orders of magnitude by the time the resonance period ends. 

\begin{figure}[htbp] 
   \centering
\includegraphics[width=.99\columnwidth]{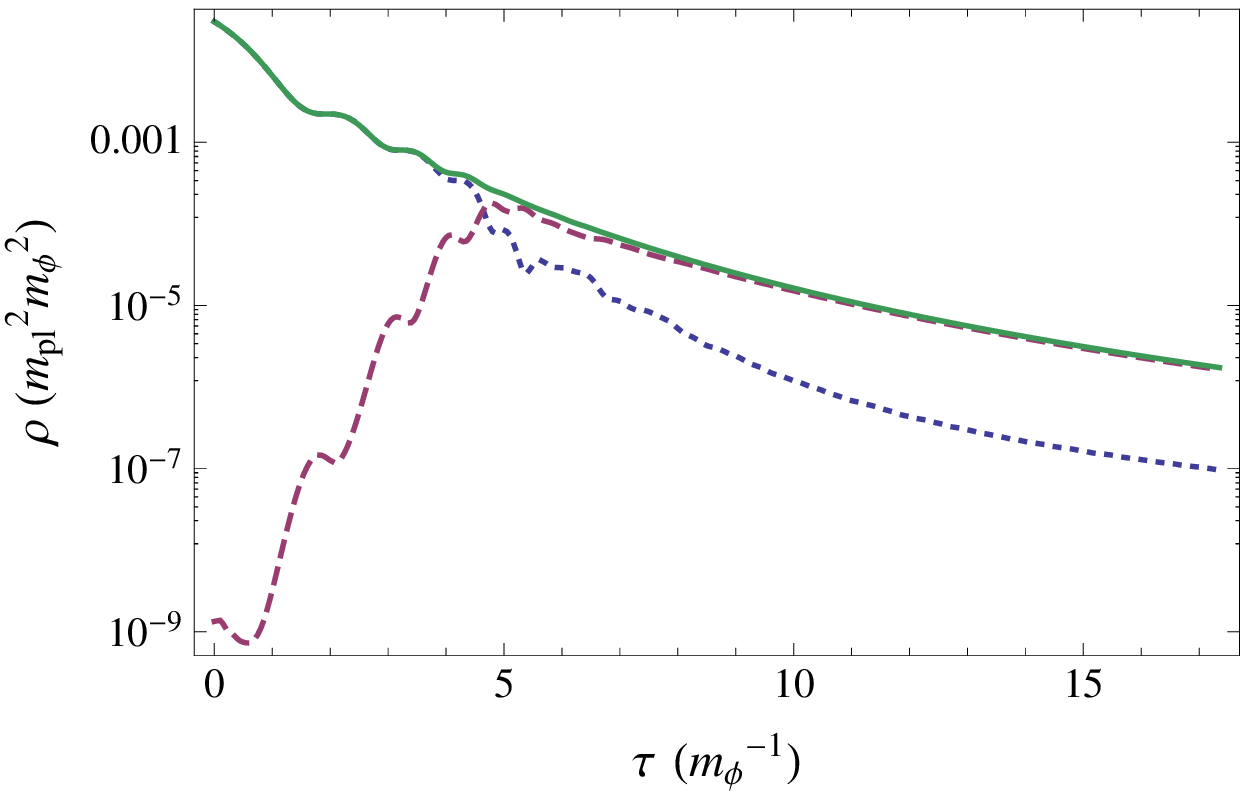}
\includegraphics[width=.99\columnwidth]{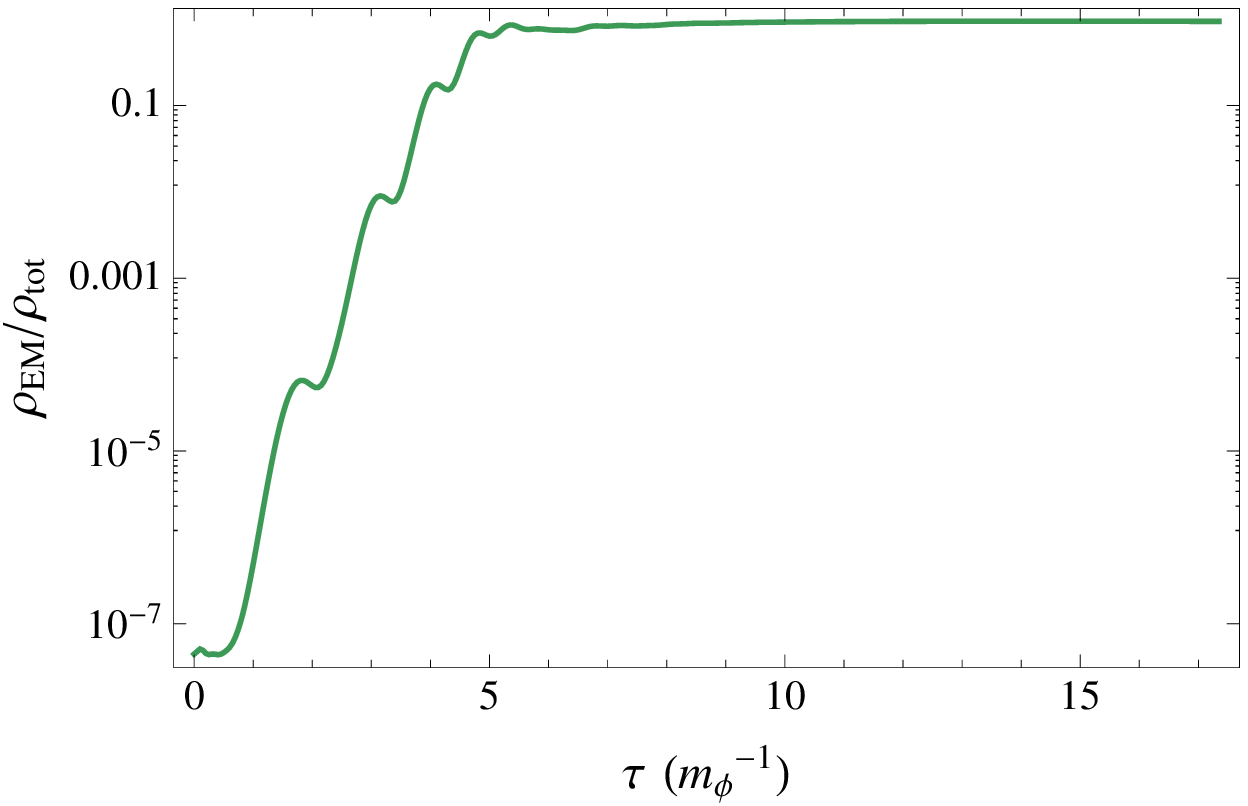}
\includegraphics[width=.99\columnwidth]{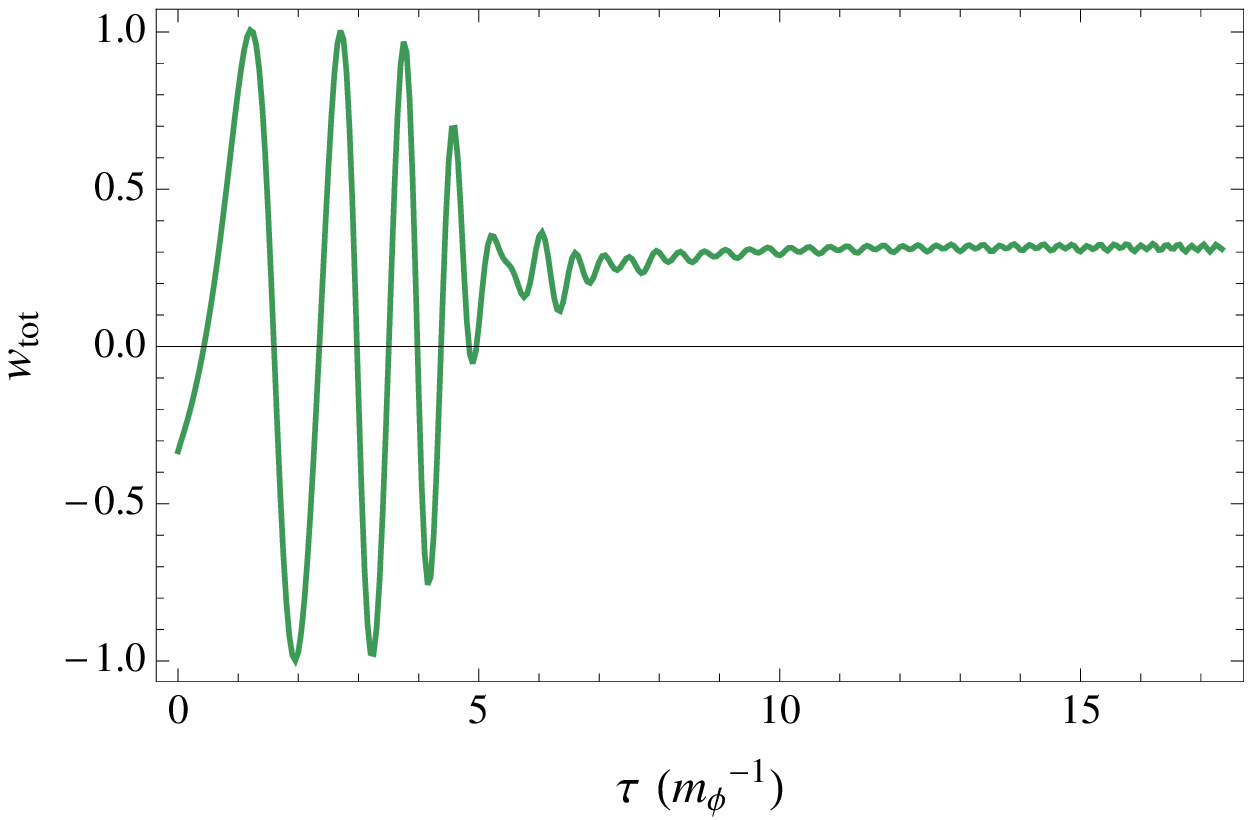}
\caption{The energy density in the inflaton (blue, dotted), the gauge fields (red, dashed), and the total (green, solid) (top). The fraction of the total energy in the gauge field (middle). The equation of state, $w=p/\rho$, during the simulation (bottom).}
   \label{fig:energies}
\end{figure}

Although the growth in variance of $A_{\mu}$ confirms the existence of a period of resonance it does not, alone, determine whether or not this resonance is sufficient to preheat the Universe. To make this assessment, we consider how much of the total energy of the simulation is transferred into the energy density of the gauge field.  We define the ratio 
\begin{equation}
\rho_{\rm EM}/\rho_{\rm tot} = \frac{\rho_{\rm EM}}{\rho_{\phi}+\rho_{\rm EM}}
\end{equation}
to parameterize the efficiency of the process. The dramatic conversion from the post-inflationary period of coherent oscillations to a radiation-dominated phase can be seen in Fig.~\ref{fig:energies}.  Here, we see the comparison of the energy in the inflaton sector, the gauge sector and the ratio of the two.  Fig.~\ref{fig:energies} also shows how the equation of state changes from an oscillating $w$, consistent with a coherently oscillating scalar field, to $w\approx1/3$.
The energy density ratio grows rapidly during the time of resonance, reaching a maximum value of $\rho_{\rm EM}/\rho_{tot}=0.95$. After the end of resonance, $\tau \sim 5\, m_\phi^{-1}$, the fields are essentially decoupled and the energy that was transferred to the gauge field remains in the gauge field.  Even though some $5$\% of the total energy is still in the $\phi$ sector, because some of the $\phi$ modes that remain populated are ultra-relativistic, with $k \gg m_{\phi}$, the total energy density remains radiation dominated. 

\begin{figure}[htbp] 
   \centering
\includegraphics[width=.99\columnwidth]{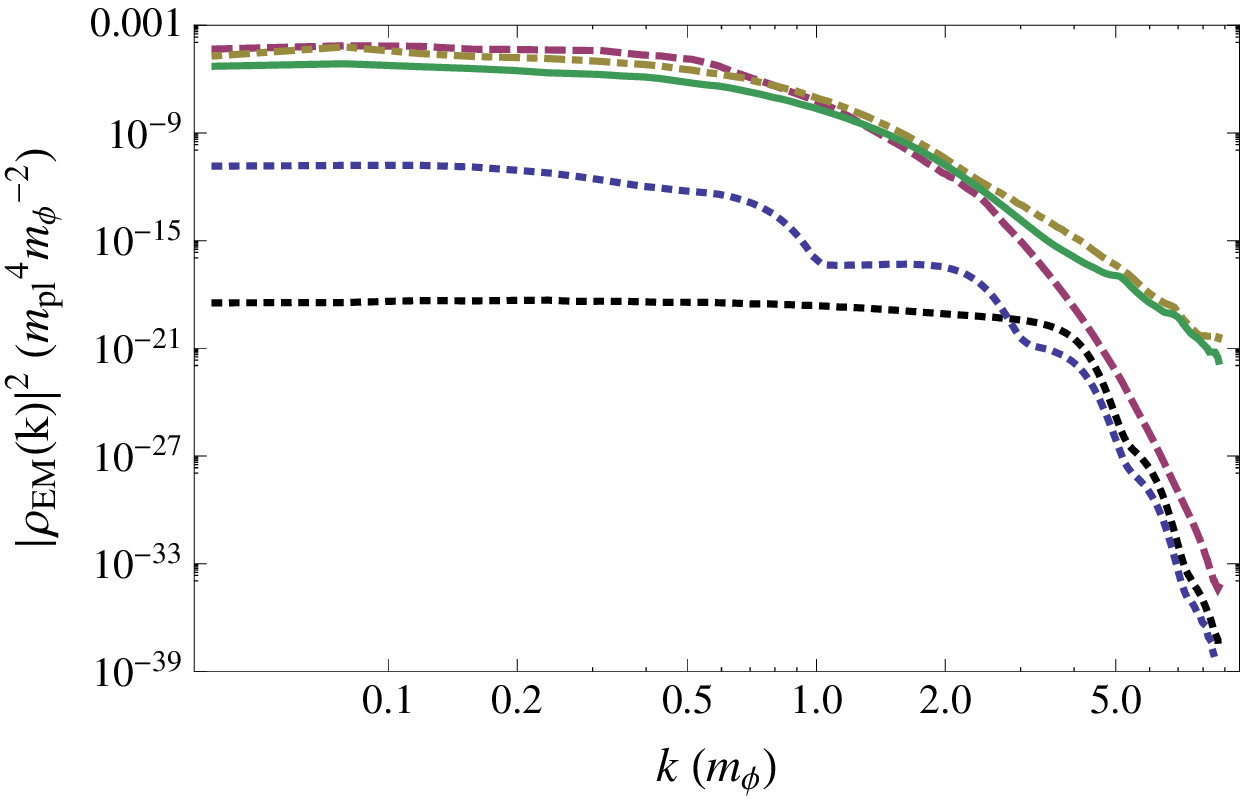}
\includegraphics[width=.99\columnwidth]{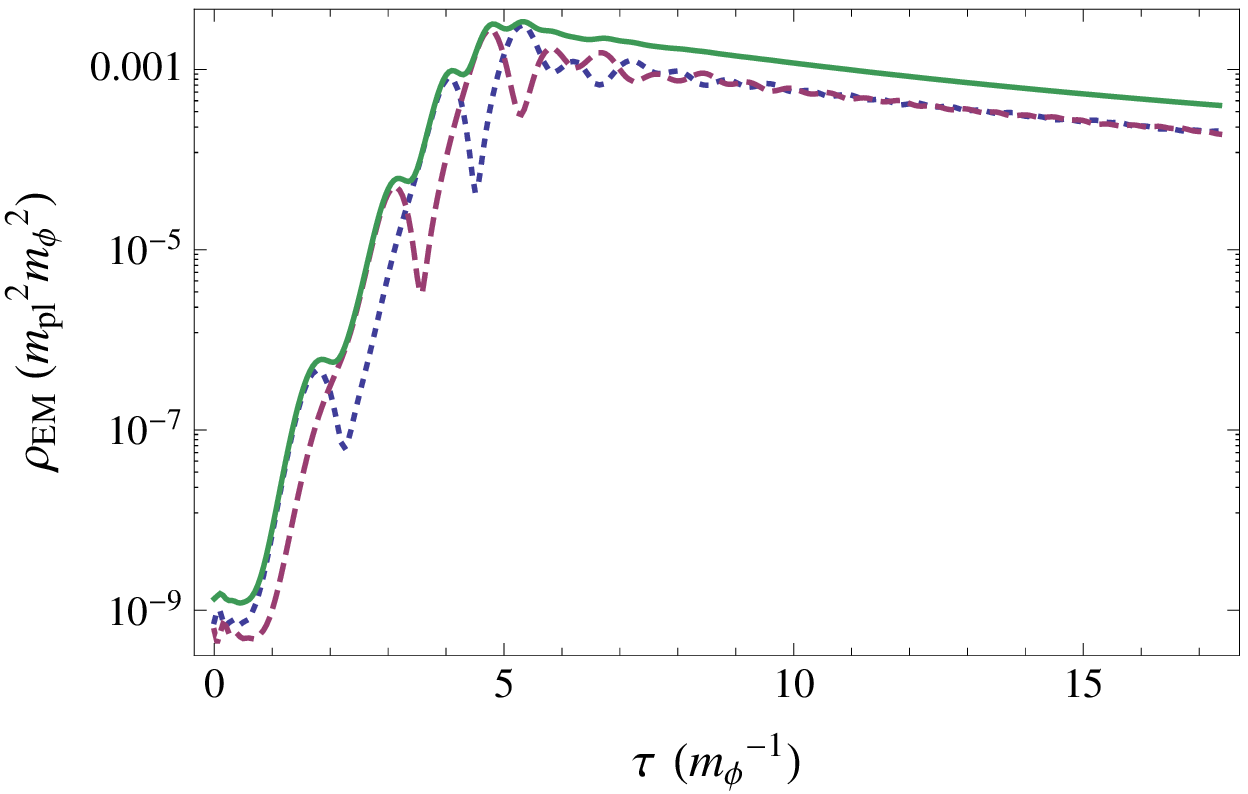}
\caption{The power spectrum of energy density in the gauge field at $\tau = 0$ (black, dotted), $\tau=2\,m_{\phi}^{-1}$ (blue, dotted), $\tau=5\,m_{\phi}^{-1}$ (red, dashed), $\tau=10\,m_{\phi}^{-1}$ (gold, dot-dashed) and at $\tau=15\,m_{\phi}^{-1}$ (green, solid) (top).  The energy density in the electric field, $\rho_{\rm E}$ (blue, dotted), the magnetic field, $\rho_{\rm M}$ (red, dashed), and the total electromagnetic energy, $\rho_{\rm EM}$ (green, solid) (bottom). }
   \label{fig:energyemsplit}
\end{figure}

We can also look at the distribution of energy in the gauge sector.  Fig.~\ref{fig:energyemsplit} shows how the amplification of electromagnetic energy occurs over time using the power spectrum, $\left|\rho_{\rm EM}(k)\right|^2$; resonance occurs broadly across the lower-frequency bands during the early stages of resonance, and higher-frequency modes are amplified toward the end of resonance.  We also see that after $\tau\sim 10\, m_{\phi}^{-1}$, the power spectrum, $\left|\rho_{\rm EM}(k)\right|^2$, changes very little.  In addition we can compare the energy in the electric field, 
\begin{equation}
\rho_{\rm E} = \frac{W}{2a^{2}}\left|\partial_\tau \vec{A} - \vec{\nabla} A_\tau\right|^2, 
\end{equation}
to the energy in the magnetic field,
\begin{equation}
\rho_{\rm M} =  \frac{W}{2a^{2}}\left| \vec{\nabla}\times \vec{A}\right|^2,
\end{equation}
as in Fig.~\ref{fig:energyemsplit}.  During the resonance period the electric and magnetic fields are amplified at slightly different times, but by the end of the simulation the energy is split evenly between the two contributions.

\begin{figure}[htb] 
   \centering
\includegraphics[width=0.99\columnwidth]{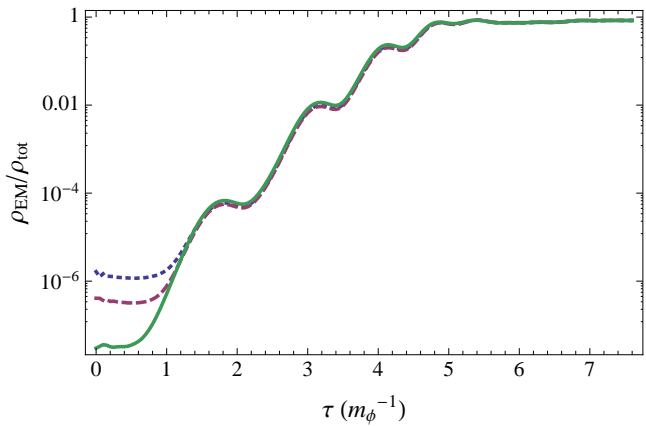}
   \caption{The energy density ratio $\rho_{\rm EM}/\rho_{tot}$ without a cutoff (blue, dotted), $k_{*}=k_{nq}/2$ (red, dashed) $k_{*}= k_{nq}/4$ (green, solid). In these simulations, $m_{\phi}=10^{-6}\,m_{\rm pl}$ and $M=0.08\,m_{\rm pl}$.}
   \label{fig:cutoffedratio}
\end{figure}

At this point we can look for any dependence that our choice of cutoff, $k_*$ might have on this outcome.  In Fig.~\ref{fig:cutoffedratio} we see the energy density ratio $\rho_{\rm EM}/\rho_{tot}$, over time for various choices of the cutoff, $k_{*}$. In other words, the scales that play a role in preheating are well resolved by our simulations and the final state of the preheated Universe is the same.

\begin{figure}[htbp] 
\centering
\includegraphics[width=0.99\columnwidth]{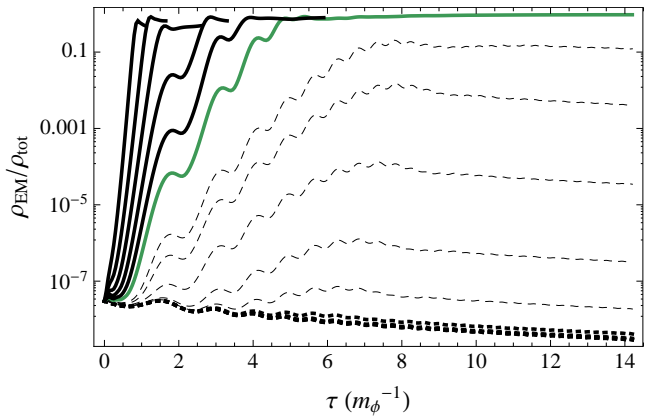}
\caption{Plot of the energy density ratio $\rho_{\rm EM}/\rho_{tot}$ where $m_{\phi}=10^{-6}\,m_{\rm pl}$ and for $M>0.04\,m_{\rm pl}$ (dotted lines), $0.017\,m_{\rm pl}<M<0.04\,m_{\rm pl}$ (dashed lines),  and for $M<0.017\,m_{\rm pl}$ (solid lines). The green line is $M=0.016\,m_{\rm pl}$.}
\label{fig:multed}
\end{figure}

Next we can quantify how the interaction parameter $M$ affects preheating. We explore values of $M$ between $0.005\,m_{\rm pl} \lesssim M \lesssim \,m_{\rm pl}$.  Fig.~\ref{fig:multed} shows the energy density ratio, $\rho_{\rm EM}/\rho_{\rm tot}$ over the simulation time. The smaller $M$ values have faster growth of the energy density ratio $\rho_{\rm EM}/\rho_{\rm tot}$. This plot shows us that larger $M$ values correspond to a more efficient (and faster) resonance period. We note that runs with very high $M\gtrsim 0.04\,m_{\rm pl}$ terminate early since the resonance in the gauge field is so broad that power builds up in high-frequency modes and destabilizes the gauge field evolution.  For the largest values of $M$ that we test, resonance occurs quickly and efficiently before the end of the first full oscillation of the inflation.
Marginal values of $0.017\,m_{\rm pl}< M< 0.04\,m_{\rm pl}$ correspond to incomplete resonance, where the gauge field does not acquire enough energy from the inflaton before the resonance mechanism shuts off.

\begin{figure}[htbp]
\centering
\includegraphics[width=\columnwidth]{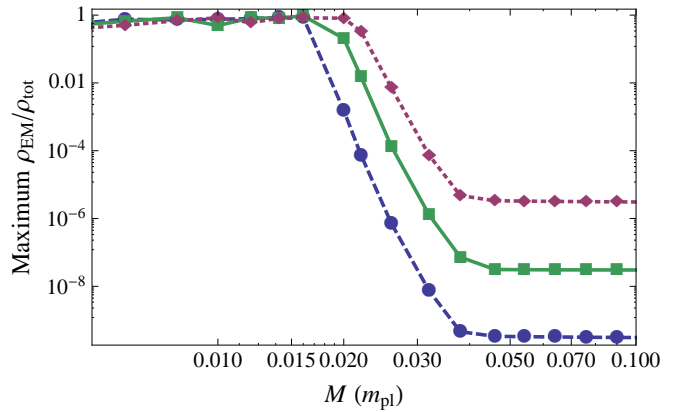}
\caption{Plot of the maximum value of $\rho_{\rm EM}/\rho_{tot}$ vs. $M$ for $m_{\phi}=10^{-5}m_{\rm pl}$ (dotted red line with diamonds), $m_{\phi}=10^{-6}m_{\rm pl}$ (solid green line with squares), and $m_{\phi}=10^{-7}m_{\rm pl}$ (dashed blue line with circles).}
\label{fig:moneyplot}
\end{figure}

We can further probe the effectiveness of preheating on the parameter $M$ by looking at the maximum value of the energy density ratio in Fig~\ref{fig:moneyplot}. The green squares in Fig.~\ref{fig:moneyplot} show the maximum value of the ratio $\rho_{\rm EM}/\rho_{tot}$ for different simulations, each with a different value of $M$ with $\phi_{0}=0.2\,m_{\rm pl}$ and $m_{\phi}=10^{-6}\,m_{\rm pl}$. These points also correspond to the energy density curves in Fig.~\ref{fig:multed}.   For large values of $M$, the resonance period is not as effective as it is for small $M$;  the dramatic fall off around $M\sim0.02\,m_{\rm pl}-0.03\,m_{\rm pl}$ is a consequence of the fact that the amplitude of the field during its first oscillation is $\phi\sim 0.05\,m_{\rm pl}$, which can be seen in Fig.~\ref{fig:fields}. It is for this reason that when $M$ is much greater than $0.05\,m_{\rm pl}$, we do not see any significant resonance.  We always see strong resonant effects for $M \lesssim 0.02 \,m_{\rm pl}$, because the coupling term is relevant for a significant portion of the period of coherent oscillations in $\phi$. When $M$ is on the same order as $\phi_0$, we see partial resonance.

\begin{figure}[htbp]
\centering
\includegraphics[width=\columnwidth]{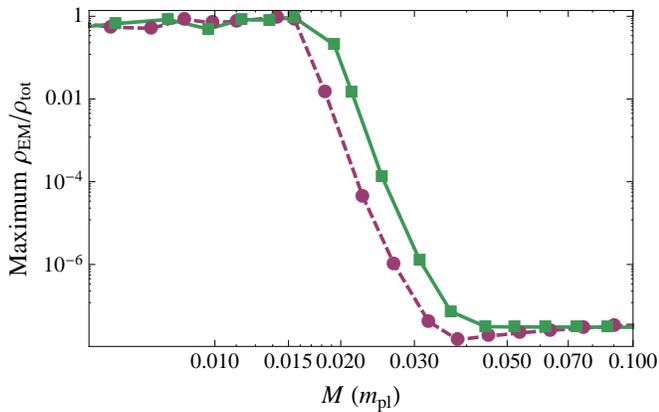}
\caption{Plot of the maximum value of $\rho_{\rm EM}/\rho_{tot}$ vs. $M$ for $\phi_0=0.2\,m_{\rm pl}$ (green line with squares) and $\phi_0 = 0.141\, m_{\rm pl}$ (dashed red line with circles), where $m_{\phi}=10^{-6}\,m_{\rm pl}$. Note that the green curve has the same set of parameters as the green curve in Fig.~\ref{fig:moneyplot}.}
\label{fig:moneyplotvary0}
\end{figure}

The effect of changing $m_\phi$ can also be seen in Fig.~\ref{fig:moneyplot}.  We see that, for small values of $M$, the final state of the simulation is a radiation-dominated, preheated Universe, independent of $m_\phi$.  For larger values of $M$, the peak value of the fraction $\rho_{\rm EM}/\rho_{\rm tot}$ depends on $m_\phi$; of course this maximum value corresponds to the initial fraction of each simulation, and is a function of the initial conditions. The dramatic difference between these states---the range of parameters for which resonance is partial---again occurs near $M\sim0.02\,m_{\rm pl} -0.03\,m_{\rm pl}$.  We can change this location by changing $\phi_0$, though compatibility with successful inflation dictates that this change should not be too large. Fig.~\ref{fig:moneyplotvary0} shows the effect of this change:  a lower initial amplitude for $\phi_0$ moves the drop off to smaller values of $M$. Considering that the coupling function depends on the ratio $\phi/M$, then a change of $\sqrt{2}$ between the initial values of $\phi_0$ in the curves illustrated in  Fig.~\ref{fig:moneyplotvary0} agrees with a shift of the drop off in $M$ by the same factor of $\sqrt{2}$.
 
\section{Discussion}

We have presented the first high-resolution finite-time lattice simulations of preheating for a $U(1)$ abelian gauge field coupled to a scalar field. The ability to simulate coupled scalar-gauge theories on an expanding background is particularly timely and important, since the models and dynamics of reheating are moving past toy models and the focus is moving toward understanding how the inflaton can couple to the fields in some extension of the Standard Model. 

Here we have shown that an abelian $U(1)$, massless, electromagnetic field can be used as a mechanism by which energy can be moved from the inflaton into particles.  The process of gauge field preheating, ``gauge-preheating," presents two significant improvements to the standard preheating scenario: (1) it is likely that gauge fields exist in whatever extension of the Standard Model, and that the symmetries of these gauge fields have a $U(1)$ subgroup and (2) it does not require additional channels of decay, or tachyonic instabilities, in order to generate the radiation-dominated Universe that is needed for primordial nucleosynthesis.  In our model once the energy is moved to the gauge field during resonance there is no movement of energy back to the scalar field that one would normally see with a coupling between two scalar fields \cite{Dufaux:2006ee}. The model that we have proposed, however, has a natural termination; once $\phi$ becomes small, the coupling term becomes negligible. This ensures that whatever energy is deposited into the gauge field will remain there. Thus our model gives a method for creating a Universe at the end of inflation with a majority of its energy in a gauge field $A_{\mu}$.  Furthermore, since at late times the conformal coupling constant is close to unity, the standard dynamics of a gauge field, electromagnetism, is recovered.

In traditional preheating, a significant fraction of the energy of the inflaton is transferred into a massless degree of freedom.  In preheating scenarios, optimistic predictions say that the coupled degree of freedom will have approximately the same energy as the inflaton \cite{Battefeld:2009xw}, even with three leg interactions or tachyonic instabilities.  These scenarios then rely on additional mechanisms or interactions to deplete the remaining energy in the inflation.  Here, we see that the fraction of energy transferred into the gauge field is much more substantial; the energy in the inflaton is depleted dramatically.  For the example of Section~\ref{results} where $M=0.016\,m_{\rm pl}$, only a few percent of the total energy remains in the inflaton sector, whereas smaller values of $M$ are even more efficient.  Even still, the existence of any energy left in the inflaton sector has the potential to return the Universe to a matter-dominated era.  We must then rely on additional couplings to additional degrees of freedom to fully diminish the energy in the inflaton.

The complexity of the equations of motion of our theory necessitated the creation of a new code, GABE, to evolve the coupled partial differential equations.  This software also represents the potential to couple the inflaton to more sophisticated gauge fields, and opens the door to a study of preheating in other models of inflation, such as Chromo-Natural Inflation \cite{Adshead:2012kp}.

\section{Acknowledgments}  
We want to thank David Kaiser, Alan Guth and Ruth Durrer for conversations that inspired this work.  We also want to thank Eugene Lim for years of conversations about preheating and suggesting the name ``Gauge-Preheating." We also want to thank Alexander New for insight on how to treat the gauge field in this model.  The work of RRC is supported in part by NSF PHY-1068027 at Dartmouth. 
 JTD and JTG are supported by the National Science Foundation, PHY-1068080, and a Cottrell College Science Award from the Research Corporation for Science Advancement.  


\end{document}